\documentclass{article}

\usepackage{PRIMEarxiv}

\usepackage[utf8]{inputenc} 
\usepackage[T1]{fontenc}    
\usepackage{hyperref}       
\usepackage{url}            
\usepackage{booktabs}       
\usepackage{amsfonts}       
\usepackage{nicefrac}       
\usepackage{microtype}      
\usepackage{lipsum}
\usepackage{fancyhdr}       
\usepackage{graphicx}       
\usepackage{amsmath}

\usepackage[resetlabels,labeled]{multibib}

\newcites{S}{Supplemental References}

\usepackage[%
    binary-units=true,
    prefixes-as-symbols=false,
]{siunitx}

\DeclareSIUnit{\microsecond}{\SIUnitSymbolMicro s}

\title{Differentiable Folding for Nearest Neighbor Model Optimization
}

\author{
  Ryan K. Krueger \\
  School of Engineering and Applied Sciences \\
  Harvard University \\
  Cambridge, MA 02138, USA \\
  \texttt{ryan\_krueger@g.harvard.edu} \\
   \And
  Sharon Aviran \\
  Department of Biomedical Engineering and Genome Center\\
  University of California, Davis \\
  Davis, CA 95616, USA \\
  \texttt{saviran@ucdavis.edu} \\
  \And
  David H. Mathews \\
  Department of Biochemistry and Biophysics and Center for RNA Biology \\
  University of the Rochester Medical Center\\
  Rochester, NY 14642, USA \\
  \texttt{david\_mathews@urmc.rochester.edu} \\
  \And
  Jeffrey Zuber \\
  Department of Biochemistry and Biophysics and Center for RNA Biology \\
  University of the Rochester Medical Center\\
  Rochester, NY 14642, USA \\
  \texttt{jzuber@gmail.com} \\
  \And
  Max Ward \\
  Department of Computer Science and Software Engineering \\
  The University of Western Australia \\
  Perth, WA, Australia \\
  \texttt{max.ward@uwa.edu.au}
}

\begin{document}
\maketitle

\begin{abstract}
The Nearest Neighbor model is the \emph{de facto} thermodynamic model of RNA secondary structure formation and is a cornerstone of RNA structure prediction and sequence design.
The current functional form (Turner 2004) contains $\approx13,000$ underlying thermodynamic parameters, and fitting these to both experimental and structural data is computationally challenging.
Here, we leverage recent advances in \emph{differentiable folding}, a method for directly computing gradients of the RNA folding algorithms, to devise an efficient, scalable, and flexible means of parameter optimization that uses known RNA structures and thermodynamic experiments.  
Our method yields a significantly improved parameter set that outperforms existing baselines on all metrics, including an increase in the average predicted probability of ground-truth sequence-structure pairs for a single RNA family by over 23 orders of magnitude.
Our framework provides a path towards drastically improved RNA models, enabling the flexible incorporation of new experimental data, definition of novel loss terms, large training sets, and even treatment as a module in larger deep learning pipelines.  
We make available a new database, RNAometer, with experimentally-determined stabilities for small RNA model systems.
\end{abstract}


\clearpage

\section{Introduction}

The Nearest Neighbor model (NN model), a.k.a. the Turner rules, is the gold standard thermodynamic model of RNA secondary structure formation.
The model assigns a free energy by decomposing a sequence-structure pair into a non-overlapping set of ``loops'' and ascribing a free energy change to each loop.
This is amenable to dynamic programming, enabling the efficient calculation of the partition function~\cite{mccaskill1990equilibrium}.
The model and corresponding suite of algorithms undergird popular software packages such as mfold~\cite{zuker2003mfold}, NUPACK~\cite{zadeh2011nupack}, ViennaRNA~\cite{lorenz2011viennarna}, and RNAstructure~\cite{reuter2010rnastructure}.

The NN model consists of $\approx13,000$ thermodynamic parameters~\cite{Zuber2017}.
The standard fitting procedure involves linearly interpolating $\approx300$ parameters to experimentally measured free energy changes~\cite{mathews2004incorporating,zuber2018analysis}. 
The complete parameter set is then extrapolated from this set of base parameters.
This procedure is complicated; though tractable, it is challenging to reproduce, requires substantial domain expertise, cannot include known sequence-structure data, and is a noisy fit given the loss in information by assuming linear dependencies.

Prior methods attempt to improve RNA structure prediction, either with advanced NN parameter fitting schemes or via alternative modeling techniques.
In Ref. \cite{andronescu2007efficient}, the authors developed two approaches to optimize parameters for the NN model.
The first was gradient descent to optimize the ``Boltzmann Likelihood'' of known RNA structures. 
This is similar to prior probabilistic methods like CONTRAfold \cite{do2006contrafold}, but incorporates thermodynamic constraints into the optimization. 
This method proved too slow to optimize parameters. 
Instead, an iterative constrained optimization heuristic was deployed. 
A key feature of Ref. \cite{andronescu2007efficient} is that the parameters are fit both to known thermodynamic experiments and to a data set of known RNA structures. 
This helps to prevent over-fitting and ensures the parameters are interpretable.
Alternative modeling techniques include (i) deep learning and (ii) generative probabilistic models via stochastic context free grammars (SCFGs), which learn the parameters of the model rather than replacing the model.
Deep learning methods have generally struggled to generalize outside their training data \cite{szikszai2022deep,sato2023recent}, a property clear in the RNA results for CASP15 \cite{das2023assessment} and CASP16.
SCFGs similarly suffer from over-fitting \cite{rivas2012range} but can achieve robust performance with careful validation \cite{rivas2013four,lu2009maxexpect}.

This work is an evolution of Ref. \cite{andronescu2007efficient} in which we demonstrate how \emph{differentiable folding} can be adapted for efficient, flexible, and transparent NN parameter fitting.
Differentiable folding is a recently developed method for RNA design in which gradients of McCaskill's recursions~\cite{mccaskill1990equilibrium} for computing the RNA partition function can be directly computed via automatic differentiation.
Here, rather than optimizing a sequence distribution with respect to a fixed model of RNA thermodynamics, we optimize the parameters of the underlying thermodynamic model using ground truth structural and thermodynamic information defined for fixed input sequences.
The flexibility of our method also lets us incorporate thermodynamic data, like Ref. \cite{andronescu2007efficient}, but also to use even more complex loss functions. 
In essence, any continuous and differentiable loss function can be optimized. 
Our method is fast enough to enable us to probe various objective functions and to do extensive cross validation.

As a demonstration, we fit parameters to minimize diverse objective functions.
First, we define individual objective functions over different data sources (i.e. structural and thermodynamic).
Given the flexibility in the choice of objective function, we can (i) control over-fitting to individual RNA families and (ii) evaluate the trade-offs imposed by each data source by performing optimizations with varying relative weights assigned to each objective.
We also explore the role of parameter inter-dependencies by performing optimizations using both the highly constrained rules of Ref. \cite{zuber2018analysis} as well as a minimal set of symmetries across parameters.
Lastly, we perform optimizations with respect to different versions of the recursions varying in their treatment of coaxial stacks, terminal mismatches, and dangling ends, the subject of previous work~\cite{ward2017advanced,ward2019determining}.

\begin{figure*}[t]
\begin{center}
\centerline{\includegraphics[width=1.0\textwidth]{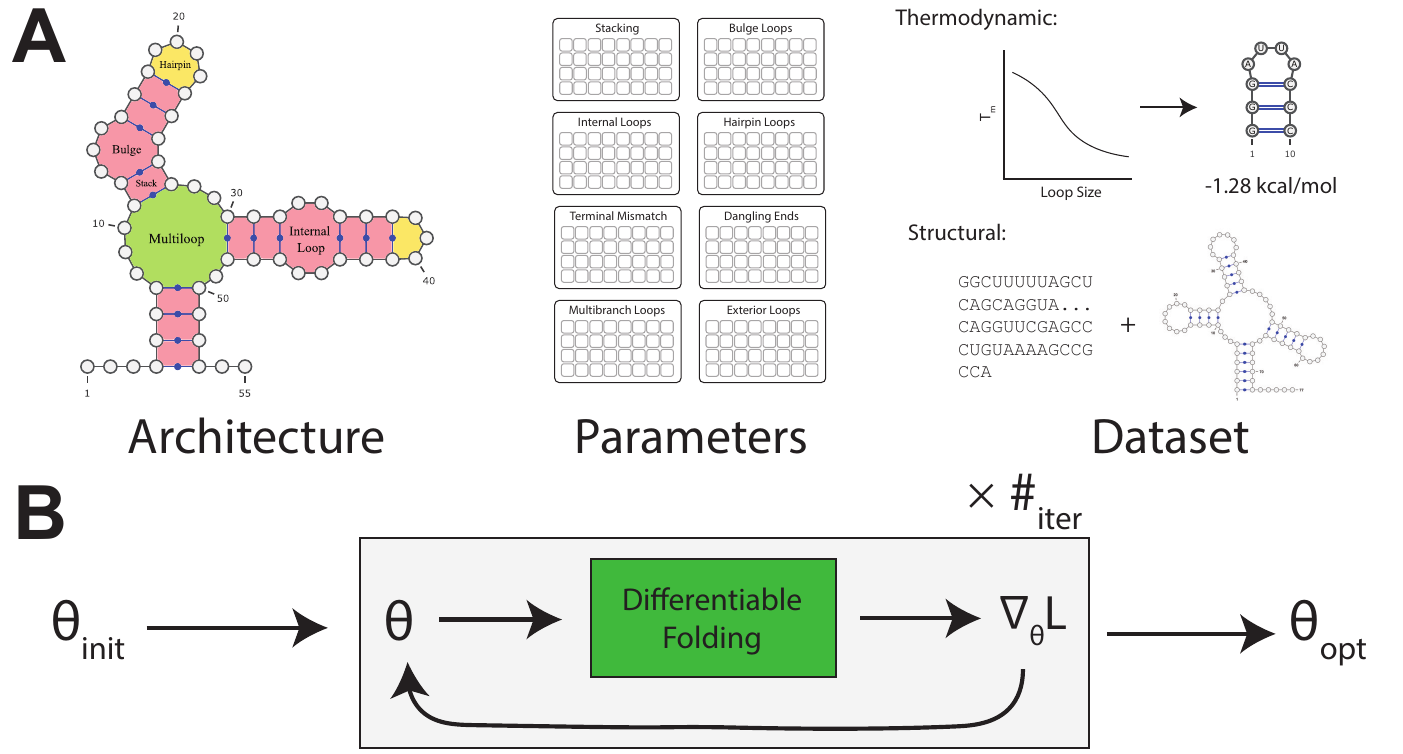}}
\caption{
An overview of our method for NN Model parameter optimization.
\textbf{A.} 
NN model parameter fitting can be formulated as an optimization problem akin to training a neural network.
The architecture is defined by the RNA folding grammar, the free parameters are (a subset of) the corresponding thermodynamic values, and the dataset comprises of (i) experimental optical melting experiments, which ascribe free energies to sequence-structure pairs, and (ii) structural data comprising of sequences and their most likely structures.
\textbf{B.} Our method for parameter optimization via differentiable folding, in which a loss function is defined over thermodynamic quantities and its gradient is computed via differentiable folding for gradient descent.
}
\label{fig:fig1}
\end{center}
\end{figure*}

Our method yields drastically improved NN parameters for both structure prediction and agreement with thermodynamic experiments. 
This is despite strict family-fold validation to prevent over-fitting.
Our optimized parameters are available via \texttt{jax-rnafold}. 

\section{Methods}


\subsection{General Purpose Framework}

In Ref. \cite{matthies2023differentiable}, a generalization of McCaskill’s algorithm is given that is well-defined over a continuous (i.e. probabilistic) sequence representation.
When implemented in an automatic differentiation framework, gradients of the partition function can be computed with respect to the sequence for inverse folding.
This paradigm is known as \emph{differentiable folding}.
Crucially, RNA design requires a fixed parameterization of the NN model.
Formally, the partition function $Z_{q, \theta}$ is parameterized by two independent parameter sets: the (continuous or discrete) sequence $q$, and the NN parameters $\theta$. 
Matthies et al. originally developed differentiable folding to enable the automatic calculation of $\nabla_q Z_{q, \theta}$ where $q$ is represented as a continuous variable.
Note that we explicitly refer to the partition function as the primary thermodynamic quantity of interest, but in practice differentiable folding can be applied to secondary thermodynamic quantities of interest, e.g. the free energy and probability of a sequence-structure pair.

\begin{figure*}[t]
\begin{center}
\centerline{\includegraphics[width=1.0\textwidth]{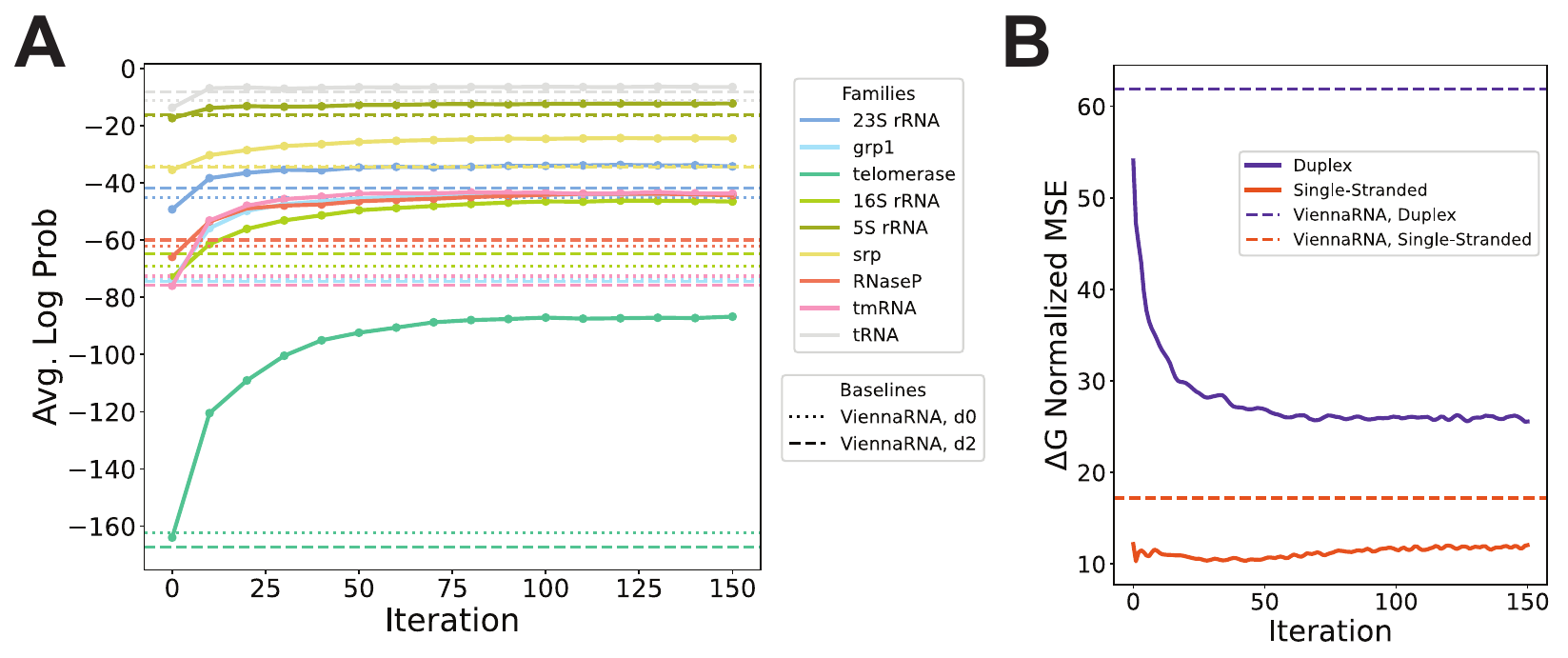}}
\caption{
Optimizing Nearest Neighbor parameters via gradient descent under our default settings (i.e. $\alpha = 0.5$, no terminal mismatches, dangling ends, or coaxial stacks (equivalent to \texttt{d0} in ViennaRNA), and the extrapolation rules of Ref. \cite{zuber2018analysis}). 
\textbf{A.} The change in the average log-probability for all sequences of length $n \leq 512$ for each family within the ArchiveII dataset. 
Since optimization is performed via stochastic gradient descent, points depict periodic evaluations of the entire dataset.
23S ribosomal RNAs are excluded from the training set.
Dashed lines depict baseline values computed via ViennaRNA with the default Turner 2004 parameters.
\textbf{B.} The change in normalized mean squared error (MSE) between the ground truth free energy values from the thermodynamic dataset of optical melting experiments and the computed values.
Dashed lines depict this value evaluated using ViennaRNA with the Turner 2004 parameters under \texttt{d0}.
}
\label{fig:fig2}
\end{center}
\end{figure*}

In this work, we adapt differentiable folding for an entirely different optimization problem: fitting the underlying NN parameters $\theta$.
Given ground-truth structural and thermodynamic data, we can define an arbitrary (continuous and differentiable) objective function $O(\theta)$ expressing the degree to which a model parameterized by $\theta$ fits the data.
We can directly compute $\nabla_{\theta}O(\theta)$ via differentiable folding and update $\theta$ via gradient descent.
This application of differentiable folding naturally scales to longer sequences than for RNA design as the RNA sequences are discrete rather than continuous, omitting the need to differentiate the memory-intensive recursions defined in Ref. \cite{matthies2023differentiable}.
In addition, our implementation is in JAX and can therefore compile to a range of targets (e.g. CPU, GPU, and TPU), rendering our method extremely efficient compared to existing work.
All optimizations were performed on a single NVIDIA 80 GB A100 GPU in less than 2 days.

\subsection{Objective Functions and Optimization Details}\label{sec:objective}

We consider two types of data: thermodynamic and structural.
Thermodynamic data consist of sequence-structure pairs with known free energy changes determined via optical melting experiments.
These thermodynamic data serve as the basis for standard NN parameter fitting schemes.
Formally, we have a library of optical melting data $M$ with $(q, s, \Delta G, \sigma^2) \in M$ where $q$ is an RNA sequence, $s$ is a valid secondary structure for $q$, $\Delta G$ is the experimentally-derived free energy change for $q$ folding into $s$, and $\sigma^2$ is the variance of $\Delta G$. 
Define the thermodynamic loss $\mathcal{L}_{\text{thermo}}$ of a given model parameterization $\theta$ as
\begin{align}
    \mathcal{L}_{\text{thermo}}(\theta) = \frac{1}{|M|}\sum_{(q, s, \Delta G, \sigma^2) \in M} \frac{(\Delta G - F_{\theta}(q, s))^2}{\sigma^2}
\end{align}
where $F_{\theta}(q, s)$ is the computed free energy of the sequence structure pair $(q, s)$ given model parameters, $\theta$.
We constructed $M$ by compiling $2,280$ optical melting experiments from $> 100$ independent publications, which we filtered to $1,817$ experiments for optimization.
We refer to this dataset as ``RNAometer'' (see Supplementary Information).

Structural data consist of sequence-structure pairs where secondary structures are determined by comparative sequence analysis. 
Formally, a library of structural data $S$ consists of pairs $(q, s^*) \in S$ where $s^*$ is the known secondary structure for sequence $q$.
In this work, we define $S$ as the ArchiveII dataset of $3,847$ RNA sequences with known secondary structures~\cite{sloma2016exact}.
ArchiveII is a standard benchmark for secondary structure prediction accuracy~\cite{mathews2019benchmark, sato2023recent, saman2022rna} and contains sequences spanning RNA families, including 16S, 23S, and 5S ribosomal RNA, group I self-splicing introns, signal recognition particle RNA, RNase P, tRNA, tmRNA, and telomerase RNA.
We preprocess all secondary structures by removing pseudoknots to leave the largest set of pseudoknot-free pairs via \texttt{RemovePseudoknots} in RNAstructure~\cite{smit2008knotted, reuter2010rnastructure}.

Following recent work on fitness functions in RNA design algorithms~\cite{ward2023fitness}, we design a structural objective function based on maximizing the probability of the structure in equilibrium,
\begin{align}
    p_{\theta}(s^* | q) = \frac{1}{Z_{q, \theta}} \exp(-\beta F_{\theta}(q, s^*))
\end{align}
where $\beta$ is the inverse of the product of temperature (set to $37 \text{ C}^{\circ}$) and the Boltzmann constant. 
This is equivalent to the ``Boltzmann Likelihood'' method of Ref. \cite{andronescu2007efficient} and is similar to how probabilistic models are often trained \cite{rivas2013four,rivas2012range,do2006contrafold}.

Care must be taken to prevent over-fitting an RNA secondary structure model to a subset of RNA families as (i) RNA families are not represented equally in structural databases and (ii) RNA families vary significantly in average sequence length and therefore in absolute scale of $p_{\theta}(s^* | q)$~\cite{szikszai2022deep, rivas2012range}.
To mitigate over-fitting, we define our structural objective function as the average of expected log-probabilities across families,
\begin{align}
    \mathcal{L}_{\text{struct}}(\theta) = \frac{1}{|\mathcal{F}|} \sum_{f \in \mathcal{F}} \frac{1}{|S_f|}\sum_{(s^*, q) \in S_f} \log \left( p_{\theta}\left(s^* | q\right) \right)
\end{align}
where $\mathcal{F}$ denotes the set of RNA families and $S_{f} \subseteq S$ is the subset of the structural dataset corresponding to family $f$.
This is equivalent to the average logarithmic geometric mean across families, as the geometric mean is equivalent to the exponential of the arithmetic mean of logarithms. 
We chose this loss function because the geometric mean is robust to differences of scale between averaged quantities. 
In this way, the optimization will not be biased towards increasing the probability of shorter sequence-structure pairs (e.g. tRNA) compared to longer ones (e.g. 16S ribosomal RNA).

Given objective functions for each data source, we express a joint objective function
\begin{align}
    \mathcal{L}(\theta) = (1-\alpha)\mathcal{L}_{\text{struct}}(\theta) + \alpha \mathcal{L}_{\text{thermo}}(\theta)
\end{align}
where $0 \leq \alpha \leq 1$ is a mixing factor which we introduce to control the relative importance of $\mathcal{L}_{\text{struct}}$ and $\mathcal{L}_{\text{thermo}}$.
Crucially, we can compute $\nabla_{\theta} \mathcal{L}(\theta)$ automatically via differentiable folding.

In practice, one does not directly optimize $\theta$ but instead a base set of parameters $\theta_{\text{base}}$ that is deterministically extrapolated to $\theta$.
This is both to preserve necessary symmetries between parameters and to mitigate over-fitting.
We consider two distinct extrapolation schemes.
First, we consider the simplest extrapolation that applies the minimal set of symmetries required to preserve thermodynamic interpretability (e.g. enforcing equal stacking parameters for stacking motifs that are identical up to $5' \to 3'$ and $3' \to 5'$ orientation).
In total, there are $12,291$ values in $\theta_{\text{base}}$ when applying these symmetries (excluding coaxial stacks).
Second, we consider a slightly modified version of the extrapolation rules introduced by Ref. \cite{zuber2018analysis} that map a set of $293$ base parameters to the full set of NN parameters.
Our modified rule set considers a set of $284$ base parameters, which excludes coaxial stacking parameters.

The final detail of the optimization problem is the treatment of terminal mismatches, dangling ends, and coaxial stacks.
These three accoutrements apply to multi-loops and exterior-loops.
The first and simplest model we consider does not include any of these contributions.
The second model we optimize allows a single nucleotide to contribute with all its possible favorable interactions, and is the default in ViennaRNA~\cite{lorenz2011viennarna}.
Following ViennaRNA naming conventions, we refer to these models as \texttt{d0} and \texttt{d2}, respectively.
We do not consider coaxial stacks in this work.

The direct calculation of $\nabla_{\theta}\mathcal{L}_{\text{struct}}(\theta)$ far exceeds the memory constraints of state-of-the-art GPUs given the large number of data points in the ArchiveII dataset.
This constraint can be alleviated via \emph{gradient accumulation}, by which gradients from multiple smaller mini-batches are collected before updating model weights, effectively simulating a larger batch size without increasing memory usage.
Furthermore, we employ a form of stochastic gradient descent by which we randomly sample 32 sequence-structure pairs from each family to estimate the average intra-family log probability at each iteration.
We also restrict the training set to sequences of length $n \leq 512$.
This includes independent folding domains for 16S and 23S rRNA, which are included in ArchiveII as complete structures and as structures divided into domains~\cite{mathews1999expanded}.
We omit the 23S rRNA and group II intron families from the calculation of $\mathcal{L}_{\text{struct}}(\theta)$ as each family has fewer than 32 sequences with $n \leq 512$.
This optimization yields an efficient calculation of $\nabla_{\theta}\mathcal{L}_{\text{struct}}(\theta)$, with each gradient update requiring $\sim 17.5$ minutes on a single NVIDIA A100 80 GB GPU.
By default, we perform 150 iterations of gradient descent with an Adam optimizer and a learning rate of $\eta = 0.1$.

\begin{table}[t!]
\tiny
\centering
\begin{tabular}{@{}ccccccccccc@{}}
\toprule
\textbf{}            & \multicolumn{2}{c}{\textbf{\begin{tabular}[c]{@{}c@{}}Thermodynamic Loss \\ (Normalized MSE)\end{tabular}}} & \multicolumn{8}{c}{\textbf{\begin{tabular}[c]{@{}c@{}}Structural Loss \\ (average log-probability)\end{tabular}}}                  \\ \cmidrule(l){2-11} 
\textbf{Parmeters}   & \textbf{Single Stranded}                                  & \textbf{Duplex}                                 & \textbf{16S} & \textbf{5S} & \textbf{grp1} & \textbf{RNaseP} & \textbf{srp} & \textbf{telomerase} & \textbf{tmRNA} & \textbf{tRNA} \\ \midrule
Optimized            & 12.0                                                      & 25.5                                            & -46.6        & -12.3       & -44.1         & -44.4           & -24.5        & -86.8              & -43.8          & -6.5          \\
Family-Fold Val.            & 12.0                                                      & 25.8                                            & -48.0        & -12.9       & -44.9         & -46.3           & -25.0        & -94.9              & -46.0          & -6.9          \\
Optimized (no rules) & 11.6                                                       & 22.3                                            & -44.4        & -11.4       & -42.3         & -43.4           & -23.4        & -82.6              & -42.2          & -6.6          \\
Family-Fold Val. (no rules)           & 11.5                                                      & 22.4                                            & -49.1        & -13.0       & -46.6         & -47.6           & -27.3        & -95.5              & -47.4          & -7.1          \\
Turner 2004          & 17.2                                                      & 61.9                                            & -69.3        & -16.5       & -73.4         & -62.1           & -34.3        & -162.3              & -72.6          & -11.3         \\
Andronescu 2007      & 25.5                                                      & 104.0                                           & -61.5        & -17.6       & -69.7         & -60.4           & -31.9        & -147.8              & -77.6          & -9.6          \\
Turner 1999          & 59.8                                                      & 122.1                                           & -65.0        & -17.4       & -73.4         & -62.3           & -34.5        & -166.1              & -78.1          & -11.6         \\ \bottomrule
\end{tabular}
\caption{
Performance on thermodynamic and structural datasets with our optimized parameters vs. baselines for \texttt{d0} recursions with $\alpha = 0.5$.
The structural loss is reported for all sequences of length $n \leq 512$ for each family within ArchiveII. 
All log-probabilities are reported in base $e$ (natural logarithm).
Parameters labeled ``Optimized'' were trained on all families, with and without the extrapolation rules of Ref. \cite{zuber2018analysis}.
For rows labeled ``Family-Fold Validation,'' structural losses represent the value obtained via optimization with the corresponding family excluded and thermodynamic losses represent the average value over all such optimizations (see Supplementary Information).
``Turner 2004,'' ``Andronescu 2007,'' and ``Turner 1999'' refer to the parameters from Ref. \cite{mathews2004incorporating}, Ref. \cite{andronescu2007efficient}, and Ref. \cite{mathews288expanded}, respectively.
16S and 5S refer to 16S and 5S ribosomal RNA, respectively. 
grp1 refers to group I introns.
}
\label{tab:d0-results}
\end{table}

\section{Results}\label{sec:results}

We first optimized the NN parameters under our default settings: no terminal mismatches, dangling ends, or coaxial stacks (\texttt{d0} in ViennaRNA), $\alpha = 0.5$ (equal weight of structural and thermodynamic losses), and the extrapolation rules of Ref. \cite{zuber2018analysis} (to conservatively control over-fitting).
We achieved substantially better performance on all objectives than any existing parameter set in ViennaRNA (Table \ref{tab:d0-results}, Figure \ref{fig:fig2}).
Optimized parameters improved the average probability of all sequence-structure pairs for 6 of 8 families in the training set by a factor of $1.9$ to $8.9 \times 10^{23}$ (Table \ref{tab:d0-avg-probs}).
Additionally, the normalized MSE between free energies obtained via optical melting experiments and computed values is $59.4\%$ and $74.7\%$ lower using our optimized parameters than using the default parameters in ViennaRNA for single-stranded and duplex RNAs, respectively.
Note that the initial loss values in our optimization do not equal the baseline values as we initialize with the default parameters in RNAstructure rather than those in ViennaRNA.

We evaluate our parameters on all sequence-structure pairs with $n > 512$ as well as on all 23S ribosomal RNA sequence-structure pairs with $n < 512$, revealing an average improvement in probability by a factor of $7.9 \times 10^4$ to $1.3 \times 10^{53}$ across all unseen datasets (Table \ref{tab:eval-d0-avg-probs}).
For example, despite not including any group2-family sequences in our training set, our parameters improve the average probability by a factor of $2.8 \times 10^{7}$.
As an additional measure against over-fitting, we performed family-fold validation. 
Family-fold validation is a form of cross validation where each family is held out as the validation set \cite{szikszai2022deep}. 
We found that each family improves similarly when excluded from the training set (Table \ref{tab:d0-results} and Supplementary Information).


Next, we performed optimizations with variants of the objective function and the recursions. 
We first repeated the optimization under our default settings but with a range of $\alpha$ values to explore the relative tradeoff between thermodynamic and structural loss terms.
As expected, lower/higher values of $\alpha$ yield parameters with decreased/increased agreement with structural data and increased/decreased agreement with thermodynamic data (Figure \ref{fig:fig4}A).
This highlights the flexibility of our method to accommodate a desired trade-off between data sources.
We also repeated the \texttt{d0} optimization under our default settings but using an alternative definition of the per-family structural loss that optimizes the logarithm of the average probability rather than the average log-probability (see Appendix \ref{sec:diff-obj-appendix}).
This yields improved average probabilities for all but one family, though this improvement is largely due to increases in the absolute probabilities for a small subset of individual sequence/structure pairs.
This highlights both the flexibility of our framework to accommodate different objective functions as well as the importance of a carefully crafted objective.

We next repeated the default optimization using the extrapolation scheme that only applies the minimal set of symmetries to preserve thermodynamic interpretability, increasing $|\theta_{\text{base}}|$ (and therefore the degrees of freedom) from $284$ to $12,291$.
As expected, the increased degrees of freedom yielded slightly improved performance but were not as robust to family-fold validation (see Table \ref{tab:d0-results}).

Lastly, we repeated the optimization for each choice of parameter extrapolation but with the more sophisticated treatment of terminal mismatches and dangling ends as per the \texttt{d2} option in ViennaRNA.
We achieve similar improvement as with \texttt{d0}, significantly outperforming all tested parameter sets in ViennaRNA on all objectives (see Table \ref{tab:d2-results}) and generalizing to the evaluation set (see Table \ref{tab:eval-d2}).
In general, our optimized \texttt{d2} parameters slightly outperform  their \texttt{d0} counterparts, likely due to \texttt{d2} being a more expressive grammar.
For example, when extrapolating via the rules of Ref. \cite{zuber2018analysis}, the optimized \texttt{d2} parameters provide the most accurate structure predictions for 7 of 8 families (Tables \ref{tab:d0-avg-probs} and \ref{tab:d2-avg-probs}) included in the training set and 6 of 7 unseen datasets (Tables \ref{tab:eval-d0-avg-probs} and \ref{tab:eval-d2-avg-probs}).


\section{Discussion}

Our primary contribution is an efficient, flexible, and extensible means of fitting NN parameters via differentiable folding.
We apply this to obtain several substantially improved parameter sets.
The strength of our method is highlighted by the optimized parameters under our default settings.
Our optimized parameters improve structure prediction across families while also improving agreement with optical melting experiments.
Our method's ability to improve tRNA structure prediction, which is prone to over fitting, alongside all other metrics highlights the power of gradient-based optimization.
Similarly, our optimized \texttt{d0} parameters significantly outperform existing parameters on all objectives, including those for the \texttt{d2} option in ViennaRNA and from Ref. \cite{andronescu2007efficient}.

Our method enables a host of future directions in model development.
Rather than using the same parameter set (e.g. Turner 2004) for both \texttt{d0} and \texttt{d2}, our method can be applied to infer optimal parameters for each grammar. We have yet to fit parameters that fully incorporate coaxial stacks, dangling ends, and terminal mismatches (i.e., \texttt{d3}).
Also, by permitting the optimization of an arbitrary (continuous and differentiable) objective function, our method can accommodate additional data sources (e.g., chemical probing data \cite{wayment2022rna}).

There is also opportunity for continued methodological development.
First, our method could be extended to fit enthalpy and entropy parameters rather than fitting free energies directly.
This could improve the model's thermodynamic interpretability and accuracy at a wider temperature range~\cite{lu2006enthalpy}.
Second, the vanilla form of stochastic gradient descent employed in this work could be supplemented with standard optimization tools from machine learning such as overparameterization with a neural network, and conflict-free gradient updates for multi-task objectives.
Third, our method may similarly serve as a module in larger deep learning methods for RNA structure prediction.
For example, outputs of differentiable folding may serve as input to a larger neural network, and effective NN parameters may be learned simultaneously with network weights.

\subsubsection*{Software and Data}

We use the \texttt{jax-rnafold} package for differentiable folding and our parameter optimization pipeline will be made available at \url{https://github.com/rkruegs123/jax-rnafold}.
Optimized nearest neighbor parameters will also be made available in \texttt{jax-rnafold}.
The RNAometer database is made available via Zenodo: \url{https://doi.org/10.5281/zenodo.15009794}.

\section*{Acknowledgments}

R.K.K. thanks Michael P. Brenner for his mentorship and support, and is supported by the National Science Foundation under Grant No. UWSC13223.
S.A. is supported by NIH Grant No. R21GM148835.
D. H. M. is supported by NIH Grant No. R35GM145283.


\bibliographystyle{unsrt}  
\bibliography{main}

\clearpage

\renewcommand{\thefigure}{S\arabic{figure}}
\setcounter{figure}{0}  

\renewcommand{\thetable}{S\arabic{table}}
\setcounter{table}{0}  

\appendix

\section{RNAometer Thermodynamic Dataset}

One contribution of this work that enables our formulation of the optimization problem is a database of optical melting experiments for the determination of nearest neighbor parameters.
Optical melting experiments are a standard means of measuring thermodynamic parameters for nucleic acids by monitoring the absorbance of a nucleic acid sample as it is heated, allowing the determination of the temperature at which the nucleic acid transitions from a one structural state to another~\citeS{schroeder2009optical}. 
This data commonly serves the basis of nearest neighbor parameter fitting.

We compiled 2280 optical melting experiments from $>100$ independent publications. 
We then filtered this set to 1817 experiments that were used in this work as the experimental dataset. 
These were the subset of experiments that were performed in 1 M Na\textsuperscript{+} and were consistent with two-state transitions by comparison of curve fit methods~\citeS{andronescu2014determination}.
These experiments were specifically curated to include a diverse representation of nearest neighbor loop motifs, including both single-stranded ($n=383$) and duplex ($n=1,434$) RNAs.

We refer to this database as RNAometer and make it publicly available via Zenodo: \url{https://doi.org/10.5281/zenodo.15009795}. 
We explicitly tabulate the publications from which each experiment originates.
The set of publications considered is as follows:
~\citeS{shu1999isolation, proctor2002isolation, kierzek1986polymer, carter2008thermodynamic, strom2015thermodynamic, laing1996model, ziomek2002thermal, diamond2001thermodynamics, hall1991thermodynamic, clanton20083, nguyen2010consecutive, phan2017advancing, giese1998stability, dale2000test, znosko2002thermodynamic, vecenie2004stability, o2005stability, vecenie2006sequence, o2006comprehensive, blose2007non, miller2008thermodynamic, mccann2011non, chen2012testing, lim2012stability, kent2014non, serra1997improved, nelson1981dna, antao1991thermodynamic, antao1992thermodynamic, petersheim1983base, hickey1985effects, freier1985improved, freier1986free, freier1986improved, sugimoto1987sequence, longfellow1990thermodynamic, he1991nearest, santalucia1991functional, serra1993rna, walter1994stability, wu1995periodic, mcdowell1996investigation, xia1997thermodynamics, xia1998thermodynamic, kierzek1999thermodynamics, schroeder2000factors, schroeder2001thermodynamic, mathews2002experimentally, schroeder2003thermodynamic, chen2004factors, chen2006consecutive, chen2009ca+, liu2011fluorescence, berger2018surprising, groebe1988characterization, groebe1989thermal, serra2002effects, serra2004pronouced, davis2007thermodynamic, davis2008thermodynamic, christiansen2009thermodynamic, davis2010positional, sheehy2010thermodynamic, vanegas2012effects, murray2014improved, tomcho2015improved,hausmann2012,christiansen2008,badhwar2007,chen2006purine,chen2005sheared,bourdelat2005,burkard2001GG,schroeder1996three,santalucia1991biochem,peritz1991,santalucia1990ga,fink1972,porschke1973,tinoco1973,borer1974,breslauer1975,martin1980,albergo1981gc3,albergo1981solvent,freier1983ggcc,hickey1985a7u7,freier1985dangle,freier1986hbond,sugimoto1986,sugimoto1987mm,tuerk1988,turner1988,walter1994pnas,walter1994coax,serra1994model,molinaro1995,williams1986,kim1996,mcdowell1997nmr,nakano1999,testa1999,ohmichi2002,znosko2002pa,shankar2007,thulasi2010,gu2013,crowther2017}

\newpage

\section{Optimization Details}

\begin{figure*}[ht]
\begin{center}
\centerline{\includegraphics[width=1.0\textwidth]{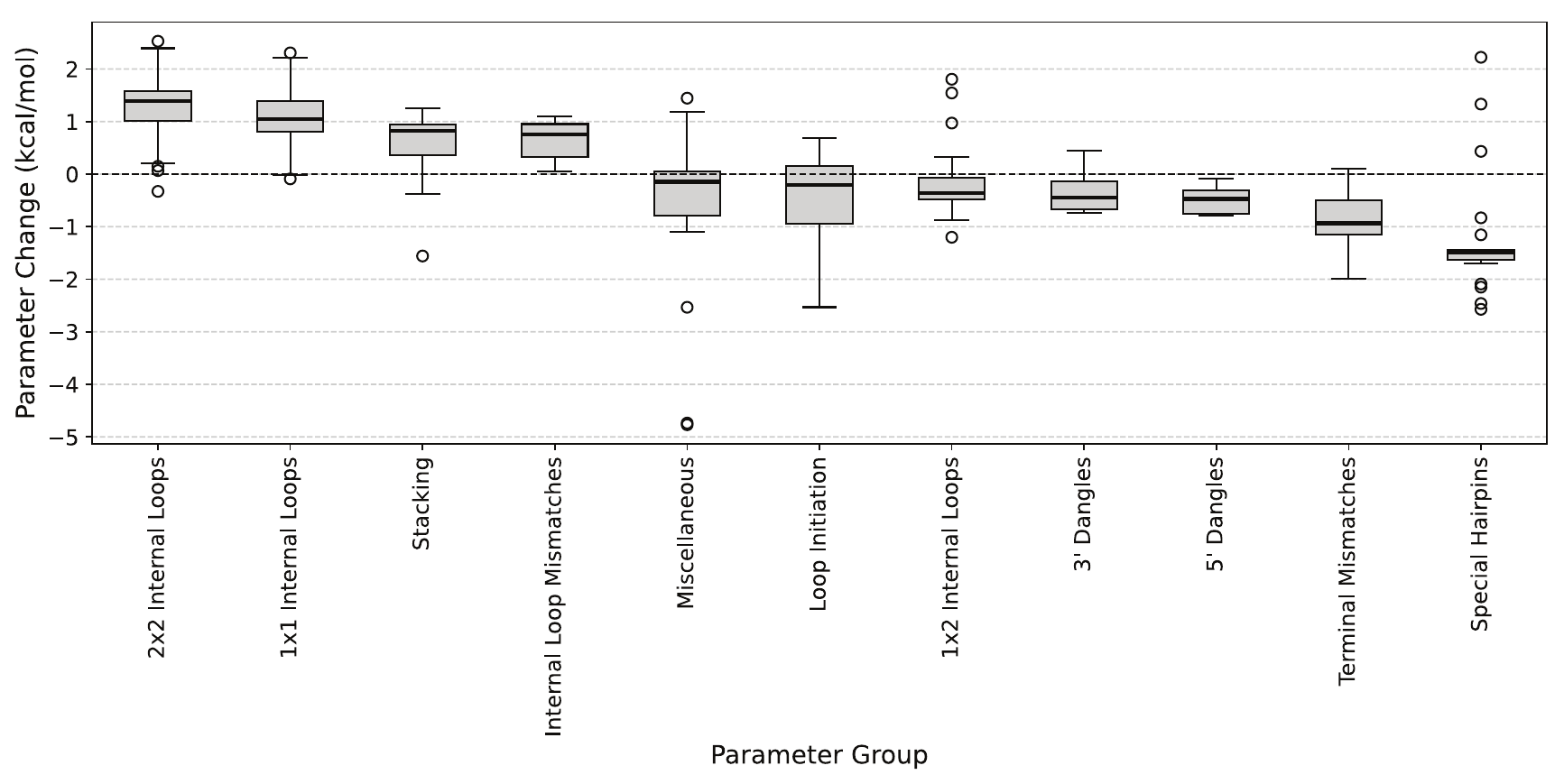}}
\caption{
Absolute changes in parameter values, grouped by parameter type, for the optimization depicted in Figure 2.
}
\label{fig:fig3}
\end{center}
\end{figure*}

In Section 3, we present optimized parameters for the \texttt{d0} recursions under the both the extrapolation rules of Ref. \citeS{zuber2018analysis_supp} as well as the base set of extrapolations described in Section 2.2.
We present the absolute parameter changes grouped by parameter type for the optimization with the extrapolation rules of Ref. \citeS{zuber2018analysis_supp} in Figure \ref{fig:fig3}.

\begin{figure*}[t]
\begin{center}
\centerline{\includegraphics[width=1.0\textwidth]{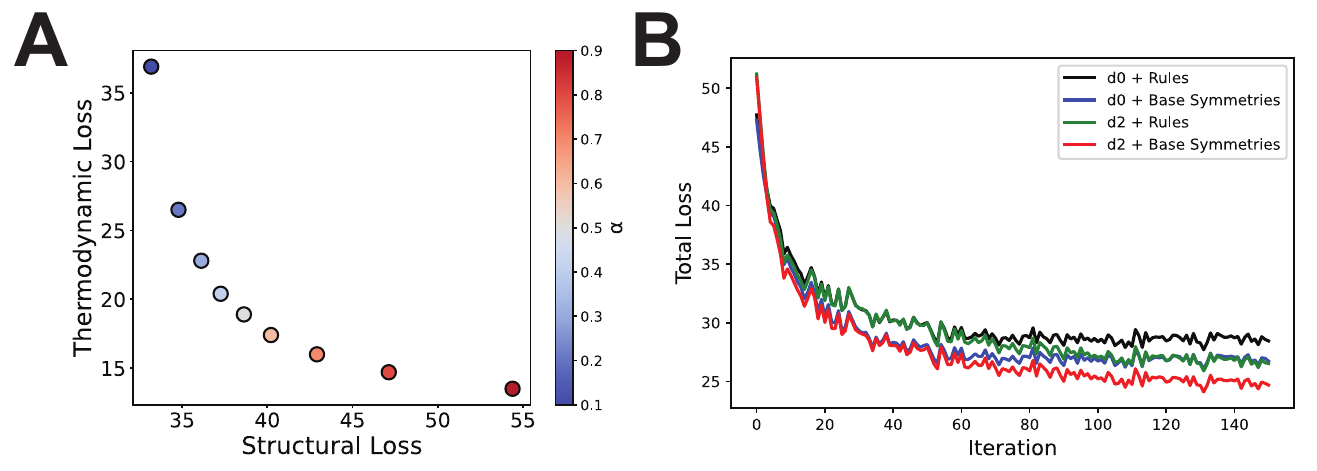}}
\caption{
Flexibly changing the formulation of the optimization problem.
\textbf{A.} The final unscaled structural and thermodynamic loss values for optimizations with the same parameters as in Figure 2 but with varying values of $\alpha$, which controls the relative importance of the two terms.
\textbf{B.} The total loss over time for four variants of the optimization problem in which we (i) follow either the \texttt{d0} or \texttt{d2} convention, and (ii) apply either the minimal set of parameter extrapolations or the more stringent extrapolation rules of Zuber et al.
}
\label{fig:fig4}
\end{center}
\end{figure*}

We also optimized parameters for different formulations of the objective function, i.e. with an extrapolation scheme applying the minimal set of symmetries, with a range of $\alpha$ values, and with the \texttt{d2} recursions (see Section 3).
We depict the tradeoff between structural and thermodynamic losses resulting from different $\alpha$ values in Figure \ref{fig:fig4}A.
We depict the total loss over time for all optimization variants with $\alpha = 0.5$ in Figure \ref{fig:fig4}B.
Lastly, we present the performance on structural and thermodynamic datasets of the optimized parameters using \texttt{d2} in Table \ref{tab:d2-results}.

\begin{table}[ht]
\tiny
\centering
\begin{tabular}{@{}ccccccccccc@{}}
\toprule
\textbf{}            & \multicolumn{2}{c}{\textbf{\begin{tabular}[c]{@{}c@{}}Thermodynamic Loss \\ (Normalized MSE)\end{tabular}}} & \multicolumn{8}{c}{\textbf{\begin{tabular}[c]{@{}c@{}}Structural Loss \\ (average log-probability)\end{tabular}}}                  \\ \cmidrule(l){2-11} 
\textbf{Parmeters}   & \textbf{Single Stranded}                                 & \textbf{Duplex}                                  & \textbf{16S} & \textbf{5S} & \textbf{grp1} & \textbf{RNaseP} & \textbf{srp} & \textbf{telomerase} & \textbf{tmRNA} & \textbf{tRNA} \\ \midrule
Optimized            & 11.2                                                     & 22.2                                             & -46.1        & -12.0       & -42.6         & -42.3           & -24.2        & -82.4              & -38.2          & -6.2          \\
Family-Fold Val.            & 11.3                                                      & 22.4                                            & -48.5        & -12.5       & -43.4         & -44.9           & -24.5        & -92.6              & -40.3          & -6.5          \\
Optimized (no rules) & 10.5                                                     & 17.5                                             & -44.2        & -10.9       & -40.2         & -41.6           & -23.3        & -78.9              & -37.1          & -6.1          \\
Family-Fold Val. (no rules)            & 11.0                                                      & 18.3                                            & -50.5        & -12.7       & -46.5         & -46.8           & -27.5        & -94.8              & -42.3          & -6.6          \\
Turner 2004          & 27.6                                                     & 87.8                                & -64.8        & -16.3       & -74.3         & -60.0           & -34.5        & -167.3              & -75.7          & -8.2          \\
Andronescu 2007      & 34.5                                                     & 135.5                                            & -60.7        & -17.7       & -76.0         & -63.9           & -34.4        & -164.3              & -85.6          & -7.8          \\
Turner 1999          & 72.4                                                     & 185.5                                            & -66.1        & -19.2       & -80.4         & -68.3           & -36.2        & -181.5              & -88.5          & -9.2          \\ \bottomrule
\end{tabular}
\caption{
Performance on thermodynamic and structural datasets with our optimized parameters vs. baselines for \texttt{d2} recursions.
Hyperparameters, parameter set and value definitions, and family descriptions are the same as in Table \ref{tab:d0-results}.
}
\label{tab:d2-results}
\end{table}

\begin{table}[ht]
\tiny
\centering
\begin{tabular}{@{}ccccccccc@{}}
\toprule
\textbf{}                   & \multicolumn{8}{c}{\textbf{Average Probability}}                                                                                                       \\ \cmidrule(l){2-9} 
\textbf{Parameters}         & \textbf{16S} & \textbf{5S} & \textbf{grp1} & \textbf{RNaseP} & \textbf{srp} & \textbf{telomerase} & \textbf{tmRNA} & \multicolumn{1}{l}{\textbf{tRNA}} \\ \midrule
Initial                     & \SI{0.00014}{}  & \SI{0.00011}{}  & \SI{7.7e-13}{}  & \SI{6.0e-11}{}  & \SI{0.016}{}  & \SI{1.8e-50}{}  & \SI{0.0013}{}  & \SI{0.0012}{}  \\
Optimized                   & \SI{0.00027}{}  & \SI{0.00055}{}  & \SI{1.0e-8}{}   & \SI{4.2e-8}{}   & \SI{0.0095}{}  & \SI{1.6e-26}{}  & \SI{0.00057}{}  & \SI{0.030}{}  \\
Family-Fold Val.            & \SI{0.00028}{}  & \SI{0.00037}{}  & \SI{8.8e-9}{}   & \SI{3.0e-8}{}   & \SI{0.0085}{}  & \SI{4.7e-29}{}  & \SI{0.00051}{}  & \SI{0.024}{}  \\
Initial (no rules)          & \SI{0.00016}{}  & \SI{0.00016}{}  & \SI{2.8e-12}{}  & \SI{1.5e-10}{}  & \SI{0.018}{}  & \SI{7.0e-50}{}  & \SI{0.0014}{}  & \SI{0.0018}{}  \\
Optimized (no rules)        & \SI{0.00066}{}  & \SI{0.00082}{}  & \SI{1.4e-8}{}   & \SI{4.5e-8}{}   & \SI{0.010}{}  & \SI{1.4e-25}{}  & \SI{0.00063}{}  & \SI{0.032}{}  \\
Family-Fold Val. (no rules) & \SI{0.00057}{}  & \SI{0.00039}{}  & \SI{1.2e-8}{}   & \SI{3.0e-8}{}   & \SI{0.0087}{}  & \SI{1.1e-29}{}  & \SI{0.00056}{}  & \SI{0.022}{}  \\
Turner 2004                 & \SI{0.00015}{}  & \SI{0.00021}{}  & \SI{8.5e-13}{}  & \SI{1.3e-10}{}  & \SI{0.018}{}  & \SI{8.4e-49}{}  & \SI{0.0014}{}  & \SI{0.0059}{}  \\
Andronescu 2007             & \SI{5.8e-6}{}  & \SI{3.7e-5}{}  & \SI{1.4e-11}{}  & \SI{9.2e-12}{}  & \SI{0.011}{}  & \SI{2.1e-45}{}  & \SI{0.00022}{}  & \SI{0.0085}{}  \\
Turner 1999                 & \SI{0.00010}{}  & \SI{0.000089}{}  & \SI{2.2e-12}{}  & \SI{1.1e-11}{}  & \SI{0.018}{}  & \SI{4.6e-50}{}  & \SI{0.00086}{}  & \SI{0.0030}{}  \\ \bottomrule
\end{tabular}
\caption{
Average probabilities per family for the \texttt{d0} optimization depicted in Figure 2. 
Hyperparameters, parameter set and value definitions, and family descriptions are the same as in Table \ref{tab:d0-results}.
}
\label{tab:d0-avg-probs}
\end{table}

\begin{table}[ht]
\tiny
\centering
\begin{tabular}{@{}ccccccccc@{}}
\toprule
\textbf{}                   & \multicolumn{8}{c}{\textbf{Average Probability}}                                                                                                       \\ \cmidrule(l){2-9} 
\textbf{Parameters}         & \textbf{16S} & \textbf{5S} & \textbf{grp1} & \textbf{RNaseP} & \textbf{srp} & \textbf{telomerase} & \textbf{tmRNA} & \multicolumn{1}{l}{\textbf{tRNA}} \\ \midrule
Initial                     & \SI{8.6e-5}{}  & \SI{0.00016}{}  & \SI{1.2e-12}{}  & \SI{2.2e-11}{}  & \SI{0.014}{}  & \SI{6.2e-49}{}  & \SI{0.0014}{}  & \SI{0.017}{}  \\
Optimized                   & \SI{0.00031}{}  & \SI{0.00081}{}  & \SI{7.2e-9}{}   & \SI{6.0e-7}{}   & \SI{0.011}{}  & \SI{8.5e-25}{}  & \SI{0.00066}{}  & \SI{0.036}{}  \\
Family-Fold Val.            & \SI{0.00030}{}  & \SI{0.00061}{}  & \SI{4.9e-9}{}   & \SI{5.7e-7}{}   & \SI{0.0098}{}  & \SI{6.0e-28}{}  & \SI{0.00063}{}  & \SI{0.027}{}  \\
Initial (no rules)          & \SI{0.00010}{}  & \SI{0.00021}{}  & \SI{2.6e-12}{}  & \SI{6.5e-11}{}  & \SI{0.016}{}  & \SI{2.6e-48}{}  & \SI{0.0014}{}  & \SI{0.021}{}  \\
Optimized (no rules)        & \SI{0.0011}{}  & \SI{0.0016}{}  & \SI{1.1e-8}{}   & \SI{1.1e-6}{}   & \SI{0.013}{}  & \SI{6.8e-24}{}  & \SI{0.00067}{}  & \SI{0.042}{}  \\
Family-Fold Val. (no rules) & \SI{0.00060}{}  & \SI{0.00067}{}  & \SI{4.5e-10}{}   & \SI{9.5e-7}{}   & \SI{0.011}{}  & \SI{2.3e-29}{}  & \SI{0.00063}{}  & \SI{0.031}{}  \\
Turner 2004                 & \SI{9.6e-5}{}  & \SI{0.00022}{}  & \SI{5.4e-13}{}  & \SI{8.4e-12}{}  & \SI{0.017}{}  & \SI{8.3e-50}{}  & \SI{0.0013}{}  & \SI{0.031}{}  \\
Andronescu 2007             & \SI{9.8e-6}{}  & \SI{2.3e-5}{}  & \SI{5.2e-13}{}  & \SI{1.8e-13}{}  & \SI{0.012}{}  & \SI{2.0e-50}{}  & \SI{0.00012}{}  & \SI{0.026}{}  \\
Turner 1999                 & \SI{3.0e-5}{}  & \SI{1.8e-5}{}  & \SI{5.5e-14}{}  & \SI{1.0e-12}{}  & \SI{0.017}{}  & \SI{2.0e-56}{}  & \SI{0.00088}{}  & \SI{0.016}{}  \\ \bottomrule
\end{tabular}
\caption{
Average probabilities per family for the \texttt{d2} optimization described in Table \ref{tab:d2-results}. 
Hyperparameters, parameter set and value definitions, and family descriptions are the same as in Table \ref{tab:d0-results}.
}
\label{tab:d2-avg-probs}
\end{table}

\clearpage

\newpage

\section{Model Evaluation}

During training, the model was fit to the full thermodynamic dataset but to only a subset of the structural dataset including sequence-structure pairs of length $n \leq 512$.
Evaluation was performed on all excluded sequence-structure pairs, including all data points with $n > 512$ and all 23S ribosomal RNA data points with $n < 512$, as the latter were too scarce to be included in training.
We report the structural prediction performance for the evaluation set for both \texttt{d0} and \texttt{d2} optimizations in Tables \ref{tab:eval-d0} and \ref{tab:eval-d2}, respectively.

In addition to this default evaluation, we also performed family-fold validation in which we excluded additional families from the training set.
For example, an optimization in which we excluded tRNA's would not include any tRNA sequence-structure pair with $n \leq 512$ in the training set, in addition to the default exclusion criteria.
As a generic evaluation of the generalizability of our fitting procedure, we perform a series of family-fold validation experiments in which we individually exclude each family that is otherwise included in the training set.
We repeat the optimization for each family in the training set so that we can evaluate the effect to which inclusion of a family in the training set results in overfitting to that family.
We report the results of these experiments in Tables 1 and \ref{tab:d2-results} where the value for a given family is calculated from a parameter set fit to a training set that excludes that family.
For example, when tRNA's are excluded from the training set in the \texttt{d0} optimization that extrapolates via the rules of Ref. \citeS{zuber2018analysis_supp}, our optimized parameters yield an average predicted log-probability of $-6.9$ across all tRNA's compared to $-6.5$ when the same sequence-structure pairs are included in the training set (see ``Optimized'' and ``Family-Fold Val.'' in Table \ref{tab:d0-results}).

\begin{table}[ht]
\centering
\begin{tabular}{@{}cccccccc@{}}
\toprule
\textbf{}            & \multicolumn{7}{c}{\textbf{Average Log. Probability}}                                                                          \\ \midrule
\textbf{Parameters}  & \textbf{16S} & \textbf{23S ($n \leq 512$)} & \textbf{23S} & \textbf{grp1} & \textbf{grp2} & \textbf{srp} & \textbf{telomerase} \\ \midrule
Optimized            & -167.5       & -34.3                       & -133.4           & -76.4         & -108.6        & -82.6        & -127.8              \\
Optimized (no rules) & -166.4       & -33.9                       & -143.4           & -75.8         & -110.5        & -82.2        & -132.6              \\
Turner 2004          & -260.0       & -45.0                       & -199.1       & -140.0        & -195.8        & -113.1       & -249.7              \\
Andronescu 2007      & -232.4       & -43.2                       & -201.6       & -132.0        & -190.2        & -118.3       & -231.8              \\
Turner 1999          & -249.2       & -45.6                       & -213.3       & -142.3        & -202.7        & -121.9       & -261.7              \\ \bottomrule
\end{tabular}
\caption{
Performance on structural dataset with our optimized parameters vs. baselines for \texttt{d0} recursions. 
Parameters were optimized via the protocol described in Table \ref{tab:d0-results}.
All families are restricted to sequences of length $n > 512$ unless otherwise specified.
16S and 23S refer to 16S and 23S ribosomal RNA, respectively.
}
\label{tab:eval-d0}
\end{table}

\begin{table}[ht]
\centering
\begin{tabular}{@{}cccccccc@{}}
\toprule
\textbf{}            & \multicolumn{7}{c}{\textbf{Average Log. Probability}}                                                                          \\ \midrule
\textbf{Parameters}  & \textbf{16S} & \textbf{23S ($n \leq 512$)} & \textbf{23S} & \textbf{grp1} & \textbf{grp2} & \textbf{srp} & \textbf{telomerase} \\ \midrule
Optimized            & -162.6       & -34.3                       & -127.8           & -71.5         & -103.3        & -81.0        & -120.2              \\
Optimized (no rules) & -162.5       & -34.0                       & -138.2           & -72.0         & -105.7        & -81.8        & -127.7              \\
Turner 2004          & -248.4       & -41.8                       & -185.5       & -148.5        & -205.1        & -112.4       & -256.5              \\
Andronescu 2007      & -235.0       & -44.9                       & -203.9       & -148.1        & -212.1        & -127.0       & -249.9              \\
Turner 1999          & -259.3       & -47.4                       & -224.1       & -161.7        & -224.3        & -129.7       & -273.9              \\ \bottomrule
\end{tabular}
\caption{
Performance on structural dataset with our optimized parameters vs. baselines for \texttt{d2} recursions. 
Parameters were optimized via the protocol described in Table \ref{tab:d2-results}.
All families are restricted to sequences of length $n > 512$ unless otherwise specified.
16S and 23S refer to 16S and 23S ribosomal RNA, respectively.
}
\label{tab:eval-d2}
\end{table}

\begin{table}[th]
\centering
\tiny
\begin{tabular}{@{}cccccccc@{}}
\toprule
\textbf{}            & \multicolumn{7}{c}{\textbf{Average Probability}}                                                                               \\ \cmidrule(l){2-8} 
\textbf{Parameters}  & \textbf{16S} & \textbf{23S ($n \leq 512$)} & \textbf{23S} & \textbf{grp1} & \textbf{grp2} & \textbf{srp} & \textbf{telomerase} \\ \midrule
Initial              & \SI{4.4e-37}{}  & \SI{1.4e-15}{}  & \SI{6.4e-31}{}  & \SI{8.9e-30}{}  & \SI{5.7e-33}{}  & \SI{8.9e-45}{}  & \SI{1.0e-108}{}  \\
Optimized            & \SI{7.4e-26}{}  & \SI{1.1e-10}{}  & \SI{1.4e-20}{}  & \SI{1.5e-21}{}  & \SI{1.6e-25}{}  & \SI{1.7e-30}{}  & \SI{1.3e-55}{}  \\
Initial (no rules)   & \SI{9.0e-36}{}  & \SI{9.5e-15}{}  & \SI{5.6e-29}{}  & \SI{4.9e-28}{}  & \SI{1.0e-30}{}  & \SI{1.5e-43}{}  & \SI{4.8e-109}{}  \\
Optimized (no rules) & \SI{6.3e-26}{}  & \SI{3.7e-10}{}  & \SI{2.7e-21}{}  & \SI{1.7e-22}{}  & \SI{3.4e-27}{}  & \SI{7.4e-30}{}  & \SI{3.4e-57}{}  \\
Turner 2004          & \SI{9.8e-35}{}  & \SI{3.6e-14}{}  & \SI{2.6e-27}{}  & \SI{1.3e-29}{}  & \SI{1.1e-31}{}  & \SI{1.3e-42}{}  & \SI{7.1e-107}{}  \\
Andronescu 2007      & \SI{1.9e-29}{}  & \SI{3.0e-12}{}  & \SI{2.3e-29}{}  & \SI{6.8e-30}{}  & \SI{2.6e-32}{}  & \SI{2.5e-45}{}  & \SI{1.1e-100}{}  \\
Turner 1999          & \SI{5.7e-33}{}  & \SI{9.2e-14}{}  & \SI{2.0e-29}{}  & \SI{1.6e-28}{}  & \SI{8.8e-30}{}  & \SI{4.6e-47}{}  & \SI{5.5e-114}{}  \\ \bottomrule
\end{tabular}
\caption{
Average probabilities per family for the \texttt{d0} optimization depicted in Figure 2.
Families are defined as in Table \ref{tab:eval-d0}.
}
\label{tab:eval-d0-avg-probs}
\end{table}

\begin{table}[th]
\centering
\tiny
\begin{tabular}{@{}cccccccc@{}}
\toprule
\textbf{}            & \multicolumn{7}{c}{\textbf{Average Probability}}                                                                               \\ \cmidrule(l){2-8} 
\textbf{Parameters}  & \textbf{16S} & \textbf{23S ($n \leq 512$)} & \textbf{23S} & \textbf{grp1} & \textbf{grp2} & \textbf{srp} & \textbf{telomerase} \\ \midrule
Initial              & \SI{6.0e-33}{}  & \SI{9.3e-14}{}  & \SI{8.5e-27}{}  & \SI{7.7e-31}{}  & \SI{5.5e-30}{}  & \SI{8.2e-42}{}  & \SI{3.8e-110}{}  \\
Optimized            & \SI{5.3e-23}{}  & \SI{1.5e-11}{}  & \SI{4.8e-20}{}  & \SI{1.9e-21}{}  & \SI{3.4e-25}{}  & \SI{2.1e-30}{}  & \SI{2.5e-52}{}  \\
Initial (no rules)   & \SI{9.2e-32}{}  & \SI{3.1e-13}{}  & \SI{3.2e-25}{}  & \SI{1.7e-29}{}  & \SI{5.0e-28}{}  & \SI{3.3e-41}{}  & \SI{1.0e-109}{}  \\
Optimized (no rules) & \SI{2.5e-22}{}  & \SI{1.0e-10}{}  & \SI{7.1e-20}{}  & \SI{1.0e-21}{}  & \SI{5.6e-26}{}  & \SI{4.6e-30}{}  & \SI{2.4e-55}{}  \\
Turner 2004          & \SI{2.3e-31}{}  & \SI{8.8e-13}{}  & \SI{8.3e-25}{}  & \SI{3.9e-32}{}  & \SI{6.7e-31}{}  & \SI{1.0e-41}{}  & \SI{4.8e-111}{}  \\
Andronescu 2007      & \SI{2.7e-28}{}  & \SI{1.0e-12}{}  & \SI{1.0e-30}{}  & \SI{8.9e-31}{}  & \SI{3.4e-36}{}  & \SI{6.6e-48}{}  & \SI{5.3e-109}{}  \\
Turner 1999          & \SI{5.3e-33}{}  & \SI{3.7e-13}{}  & \SI{1.3e-30}{}  & \SI{5.9e-33}{}  & \SI{2.6e-34}{}  & \SI{7.0e-48}{}  & \SI{1.7e-119}{}  \\ \bottomrule
\end{tabular}
\caption{
Average probabilities per family for the \texttt{d2} optimization described in Table \ref{tab:d2-results}.
Families are defined as in Table \ref{tab:eval-d0}.
}
\label{tab:eval-d2-avg-probs}
\end{table}

\clearpage

\newpage

\section{Different Objective Functions}\label{sec:diff-obj-appendix}

A major strength of our approach is the flexibility to define different objective functions.
To demonstrate this, we optimize nearest neighbor parameters using our default settings but with an alternative definition of the structural loss.
Previously, we defined the structural loss as
\begin{align}
    \mathcal{L}_{\text{struct}}(\theta) = \frac{1}{|\mathcal{F}|} \sum_{f \in \mathcal{F}} \frac{1}{|S_f|}\sum_{(s^*, q) \in S_f} \log \left( p_{\theta}\left(s^* | q\right) \right) \label{eqn:orig-struct}
\end{align}
where $\mathcal{F}$ is the set of RNA families and $S_{f} \subseteq S$ is the subset of all ground-truth sequence/structure pairs corresponding to family $f$.
Intuitively, this represents the average mean log-probability across all families.
However, there are many choices of alternative objectives to describe agreement with the structural data.
For example, consider the structural loss
\begin{align}
    \mathcal{L}_{\text{struct}}(\theta) = \frac{1}{|\mathcal{F}|} \sum_{f \in \mathcal{F}} \log \left[ \frac{1}{|S_f|}\sum_{(s^*, q) \in S_f} p_{\theta}\left(s^* | q\right) \right] \label{eqn:alt-struct}
\end{align}
that instead describes the average logarithm of the mean probability across all families.

Though only a minor algebraic change (i.e. moving the logarithm outside the inner expectation), the difference between Equations \ref{eqn:orig-struct} and \ref{eqn:alt-struct} can have profound effects on the meaning of the optimization.
Equation \ref{eqn:orig-struct} is more sensitive to low probability outliers while Equation \ref{eqn:alt-struct} is more sensitive to high probability outliers.
Specifically, optimizing with Equation \ref{eqn:alt-struct} rather than Equation \ref{eqn:orig-struct} will prioritize sequence/structure pairs for which the absolute value of the probability can be maximized substantially as optimizing such extrema will drastically improve the average probability over all sequence/structure pairs, $\frac{1}{|S_f|}\sum_{(s^*, q) \in S_f} p_{\theta}\left(s^* | q\right)$.
Alternatively, optimizing with respect to the original Equation \ref{eqn:orig-struct} will more strongly prioritize the entire data distribution rather than individual points as the expectation itself is computed over log-probabilities rather than raw probabilities.

\begin{figure*}[ht]
\begin{center}
\centerline{\includegraphics[width=1.0\textwidth]{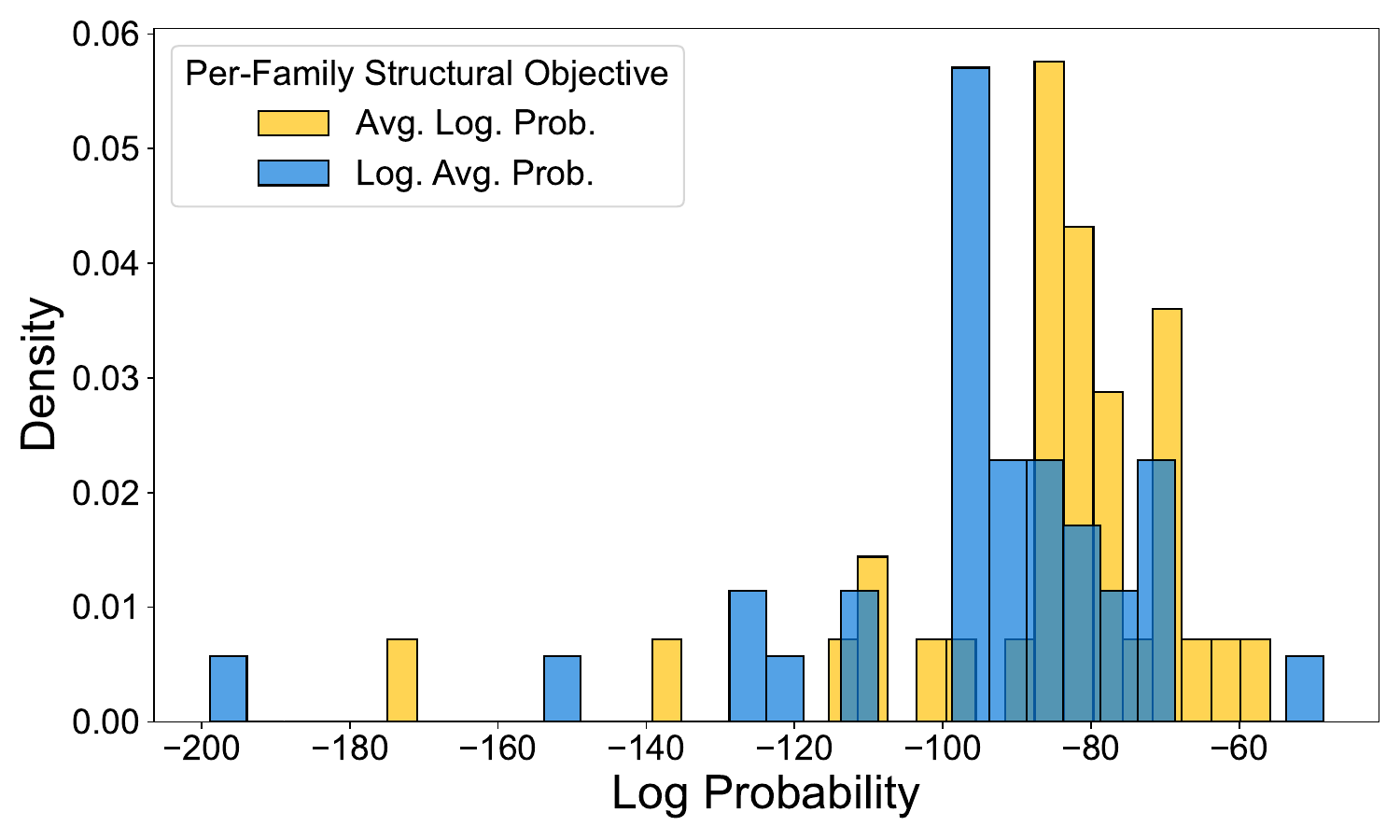}}
\caption{
The log-probabilities for all telomerase sequence/structure pairs with $n \leq 512$ under the parameters optimized using both Equation \ref{eqn:orig-struct} and Equation \ref{eqn:alt-struct} as the structural loss.
Equation \ref{eqn:orig-struct} corresponds to maximizing the average log probability per family (yellow) and Equation \ref{eqn:alt-struct} corresponds to maximizing the logarithm of the average probability per family (blue).
}
\label{fig:diff_objectives}
\end{center}
\end{figure*}

To demonstrate this, we repeated the \texttt{d0} optimization using our default parameters but using Equation \ref{eqn:alt-struct} as the structural loss instead of Equation \ref{eqn:orig-struct}.
Table \ref{tab:d0-alt-objective-avg-probs} summarizes the results of this optimization.
This optimization yielded higher average probabilities for all families but one compared to the equivalent optimization that uses Equation \ref{eqn:orig-struct} for the structural loss.
This is expected as in this case we are directly optimizing for the average probability.
However, we also find that this improvement in average probability is more strongly driven by increases in the probability of a small subset of individual sequence/structure pairs.
For example, using Equation \ref{eqn:orig-struct}, the minimum and maximum log-probabilities across all telomerase sequence/structure pairs with $n \leq 512$ are $-174.9$ and $-55.9$, respectively, while using Equation \ref{eqn:alt-struct} these values are $-198.9$ and $-48.7$ (see Figure \ref{fig:diff_objectives}).
See Table \ref{tab:d0-alt-objective-min-max} for the corresponding minimum and maximum values for all families in the training set.
This highlights how optimization under Equation \ref{eqn:alt-struct} will maximize the probability of individual sequence/structure pairs at the cost of data points with lower absolute probabilities.

\begin{table}[ht]
\scriptsize
\centering
\begin{tabular}{@{}ccccccccc@{}}
\toprule
\textbf{}                   & \multicolumn{8}{c}{\textbf{Average Probability}}                                                                                                       \\ \cmidrule(l){2-9} 
\textbf{Parameters}         & \textbf{16S} & \textbf{5S} & \textbf{grp1} & \textbf{RNaseP} & \textbf{srp} & \textbf{telomerase} & \textbf{tmRNA} & \multicolumn{1}{l}{\textbf{tRNA}} \\ \midrule
Initial                     & \SI{0.00014}{}  & \SI{0.00011}{}  & \SI{7.7e-13}{}  & \SI{6.0e-11}{}  & \SI{0.016}{}  & \SI{1.8e-50}{}  & \SI{0.0013}{}  & \SI{0.0012}{}  \\
Optimized                   & \textbf{\SI{0.0011}{}}  & \textbf{\SI{0.00071}{}}  & \SI{7.8e-09}{}   & \textbf{\SI{1.0e-07}{}}   & \textbf{\SI{0.011}{}}  & \textbf{\SI{2.1e-23}{}}  & \textbf{\SI{0.00064}{}}  & \textbf{\SI{0.031}{}}  \\
Family-Fold Val.            & \textbf{\SI{0.00041}{}}  & \textbf{\SI{0.00032}{}}  & \SI{1.4e-09}{}   & \textbf{\SI{7.5e-08}{}}   & \textbf{\SI{0.0099}{}}  & \SI{5.0e-33}{}  & \SI{0.00028}{}  & \textbf{\SI{0.030}{}}  \\ 
Turner 2004                 & \SI{0.00015}{}  & \SI{0.00021}{}  & \SI{8.5e-13}{}  & \SI{1.3e-10}{}  & \SI{0.018}{}  & \SI{8.4e-49}{}  & \SI{0.0014}{}  & \SI{0.0059}{}  \\
Andronescu 2007             & \SI{5.8e-6}{}  & \SI{3.7e-5}{}  & \SI{1.4e-11}{}  & \SI{9.2e-12}{}  & \SI{0.011}{}  & \SI{2.1e-45}{}  & \SI{0.00022}{}  & \SI{0.0085}{}  \\
Turner 1999                 & \SI{0.00010}{}  & \SI{0.000089}{}  & \SI{2.2e-12}{}  & \SI{1.1e-11}{}  & \SI{0.018}{}  & \SI{4.6e-50}{}  & \SI{0.00086}{}  & \SI{0.0030}{}  \\ \bottomrule
\end{tabular}
\caption{
Average probabilities per family for the \texttt{d0} optimization using Equation \ref{eqn:alt-struct} as the structural loss instead of Equation \ref{eqn:orig-struct}. 
Hyperparameters, parameter set and value definitions, and family descriptions are the same as in Table \ref{tab:d0-results}.
Initial values and family-wise probabilities using baseline parameter sets are identical to those in Table \ref{tab:d0-avg-probs}.
Bolded values are higher than their equivalent in Table \ref{tab:d0-avg-probs}.
}
\label{tab:d0-alt-objective-avg-probs}
\end{table}

\begin{table}[ht]
\centering
\begin{tabular}{@{}ccccc@{}}
\toprule
                    & \multicolumn{2}{c}{\textbf{Avg. Log. Prob.}} & \multicolumn{2}{c}{\textbf{Log. Avg. Prob.}} \\ \cmidrule(l){2-5} 
                    & \textbf{Min.}         & \textbf{Max.}        & \textbf{Min.}         & \textbf{Max.}        \\ \midrule
\textbf{16S}        & -133.5                & -5.1                 & \textbf{-157.5}       & \textbf{-3.1}        \\
\textbf{5S}         & -44.5                 & -3.2                 & \textbf{-52.5}        & \textbf{-2.8}        \\
\textbf{grp1}       & \textbf{-89.8}        & \textbf{-14.2}       & -89.5                 & -14.5                \\
\textbf{RNaseP}     & -131.6                & -10.9                & \textbf{-146.0}       & \textbf{-10.0}       \\
\textbf{srp}        & -150.4                & -0.5                 & \textbf{-164.8}       & \textbf{-0.4}        \\
\textbf{telomerase} & -174.9                & -55.9                & \textbf{-198.9}       & \textbf{-48.7}       \\
\textbf{tmRNA}      & -90.2                 & -1.5                 & \textbf{-96.3}        & \textbf{-1.4}        \\
\textbf{tRNA}       & -16.8                 & -0.5                 & \textbf{-20.7}        & \textbf{-0.3}        \\ \bottomrule
\end{tabular}
\caption{
Minimum and maximum log-probabilities across all sequence/structure pairs per-family under the parameters optimized using both Equation \ref{eqn:orig-struct} and Equation \ref{eqn:alt-struct} as the structural loss.
``Avg. Log. Prob.'' corresponds to Equation \ref{eqn:orig-struct} and ``Log. Avg. Prob.'' corresponds to Equation \ref{eqn:alt-struct}.
For each family, the lowest minimum and highest maximum across both optimizations are represented in bold.
}
\label{tab:d0-alt-objective-min-max}
\end{table}

\vspace*{\fill}
\clearpage

\bibliographystyleS{unsrt}
\bibliographyS{main}

\end{document}